# Memory-Efficient EDA Denoising via Knowledge Distillation for Wearable IoT Under Severe Motion Artifacts and Underwater Conditions

Yongbin Lee, Andrew Peitzsch, Youngsun Kong, Jarod Zizza, Dong-hee Kang, Farnoush Baghestani, Ki H. Chon, *Fellow, IEEE*

***Abstract*—Electrodermal activity (EDA) is widely used in wearable Internet of Medical Things (IoMT) systems for continuous health monitoring, including autonomic assessment. However, EDA signals are highly vulnerable to motion artifacts and environmental noise, limiting reliable deployment in harsh operating conditions such as underwater.**

**This study proposes a robust, deployable EDA denoising framework that generalizes across multiple measurement locations and harsh environments. The framework integrates a hybrid CNN–Transformer teacher model with a lightweight depth-wise separable CNN student model via a knowledge distillation (KD) strategy. To further improve robustness, a realistic data augmentation scheme is introduced to simulate diverse motion artifacts and environmental distortions.**

**The KD-based student model significantly reduces model size (7.87 MB to 0.51 MB) and computational cost (105.1M to 11.61M FLOPs) while maintaining denoising performance (MAE: 0.144, SNR improvement: 12.08 dB) using the public dataset validation. In real-world underwater conditions (UMAC dataset) testing, the proposed method substantially improves skin conductance response reconstruction, reducing mean absolute error from 2.809 to 0.215. Furthermore, on independent testing using the CNS-OT dataset, the denoised signals enhanced downstream CNS-OT prediction performance, achieving the highest AUROC (0.806) compared to prior denoising methods. The proposed method also improved the early prediction rate (sensitivity) from 0.550 to 0.767, enabling CNS-OT prediction up to a median of 6.9 minutes before symptom onset.**

**These results demonstrate that the proposed framework not only improves EDA signal quality but also enhances clinically relevant prediction performance while remaining suitable for deployment in resource-constrained wearable Internet of Things systems operating in harsh environments.**

Yongbin Lee, Andrew Peitzsch, Youngsun Kong, Jarod Zizza, Dong-hee Kang, Farnoush Baghestani, and Ki H. Chon are with the Department of Biomedical Engineering, University of Connecticut, Storrs, CT 06269, USA

*Corresponding author: Yongbin Lee (yongbin.lee@uconn.edu)

## I. INTRODUCTION

Electrodermal activity (EDA) measures changes in skin conductance driven by sweat gland activity under sympathetic nervous system (SNS) control [1] . Due to its non-invasive nature and strong association with autonomic arousal, EDA is widely used in wearable Internet of Medical Things (IoMT) systems for continuous health monitoring, including stress monitoring, cognitive degradation, and pain evaluation [2]–[5]. Moreover, EDA has shown potential as a supportive physiological biomarker for detecting or predicting severe conditions such as epileptic seizures and central nervous system oxygen toxicity (CNS-OT) [6]–[7]. Specifically, in naval operations, early detection or prediction of CNS-OT is crucial, as its symptoms include life-threatening conditions, such as convulsive seizures [8], and autonomic arousal reflected in EDA often precedes symptom onset [9]–[10].

Despite these advantages, many EDA studies have been conducted under controlled laboratory conditions, which may have limited applicability to real-world deployment. In practical environments, in harsh settings such as naval operations, EDA signals are highly susceptible to motion artifacts (MAs), environmental noise, and measurement location variability. These factors degrade signal quality, which compromises the reliability of physiological inference, thereby limiting the EDA-based continuous monitoring systems in safety-critical applications.

To address these issues, prior studies have explored both MA detection and denoising strategies. Early approaches focused on MA detection using methods ranging from traditional algorithms to deep learning-based classifiers [11]–[13], enabling the identification and exclusion of corrupted segments at precise time points, but not the reconstruction of EDA signals. In contrast, MA denoising approaches have relied on traditional signal processing techniques, such as wavelet-based methods including the stationary wavelet transform (SWT), to suppress noise components [14], [15]. However, these methods often struggle to preserve physiologically meaningful features, particularly when noise overlaps with the frequency band of skin conductance responses (SCRs) or under severe signal distortions such as wire interruptions (e.g., signal shearing and clipping). More recently, deep learning–based denoising approaches, such as convolutional autoencoders [16], have demonstrated improved denoising performance. Nevertheless, most existing methods primarily evaluate performance using signal reconstruction metrics (e.g., SNR, MSE) without assessing whether the reconstructed signals preserve physiological relevance or improve downstream clinical tasks.

Despite these advances, several critical challenges remain for practical deployment in wearable Internet of Things (IoT) systems. First, EDA signals exhibit substantial variability in amplitude and morphology due to differences in measurement location, environmental conditions (e.g., temperature), and subject-dependent physiological factors [17]–[20]. Second, denoising models must preserve dynamics from both tonic (baseline) and phasic (SCR) components in EDA signals to retain physiologically meaningful information. Third, wearable IoT systems impose strict constraints on memory and computational resources, requiring lightweight models suitable for embedded deployment (e.g., often < 1 MB Flash memory) [21]. Finally, the denoising methods should be evaluated not only by signal reconstruction quality but also by their impact on downstream physiological applications.

To address these challenges, we propose a robust and deployable EDA denoising framework designed for multi-site and harsh environments. The proposed approach integrates a hybrid CNN–Transformer teacher model with a lightweight depth wise separable CNN student model through a knowledge distillation (KD) strategy [22]. To further enhance robustness, we incorporate feature-wise linear modulation (FiLM) [23] to adapt to varying amplitude scales, along with a realistic data augmentation scheme that simulates diverse MAs and environmental distortions. Unlike prior work, we explicitly evaluate the physiological relevance of denoised signals by assessing improvements in both SCR reconstruction and downstream CNS-OT prediction performance.

The main contributions of this work are summarized as follows:

1) We propose a robust EDA denoising framework that generalizes across multiple measurement sites and harsh environments.
2) We introduce a teacher–student KD strategy to achieve a lightweight model deployable for resource-constrained wearable IoT devices.
3) We design a realistic data augmentation strategy to simulate real-world MAs and environmental noise.
4) We validate the proposed framework using downstream CNS-OT prediction, demonstrating improved clinical relevance beyond reconstruction metrics.

Comprehensive experiments on multi-condition datasets, including underwater environments, demonstrate that the proposed framework effectively removes real-world artifacts including dry and water immersed environments, preserves physiologically meaningful features, and enables reliable deployment in practical wearable IoT applications.

## II. System Overview

### *A. Problem Description*

As EDA reflects sweat gland activity, the fingers are commonly used as a measurement site due to their high sweat gland density. However, EDA collection on the fingers has practical limitations, as the fingers are widely used during daily activities, making long-term sensing impractical. To address this limitation, alternative measurement sites such as the foot have been explored [17], particularly for demanding applications requiring hands-free, motion-robust monitoring, such as naval operations. However, developing generalized EDA denoising models remains challenging, as EDA signals exhibit substantial variability in baseline levels and SCR characteristics across measurement sites and environmental conditions. This variability is further complicated by intrinsic physiological factors, such as asymmetric patterns across the body [20].

Fig. 1 shows representative EDA signals collected simultaneously from different body locations in underwater and in-air conditions, illustrating substantial variations in amplitude scale and SCR characteristics across measurement sites. The right foot (RF, underwater) exhibits a significantly lower baseline level and different SCR characteristics compared to the left foot (LF, in air) and right hand (RH, in air), reflecting both environmental effects and pronounced variability across locations. In particular, the underwater condition may reduce skin conductance due to lower body temperature (e.g., reduced sweat gland activity), thereby diminishing baseline levels and SCR amplitudes [19].

In addition to amplitude differences, the SCR timing and response patterns are not consistent across locations. For example, around 970–980 s, prominent SCR peaks are observed in RF and RH signals but are absent in LF, whereas around 1120–1150 s, distinct SCR activity appears only in RH. These observations indicate that even under simultaneous recording, EDA signals from different locations can exhibit heterogeneous temporal dynamics.

Such variability highlights a fundamental challenge in training robust EDA denoising models: signals collected from different locations do not necessarily share consistent reference patterns, even under similar conditions or induced artifacts. Consequently, models trained on location-specific data and constrained environmental conditions may fail to generalize across diverse real-world scenarios.

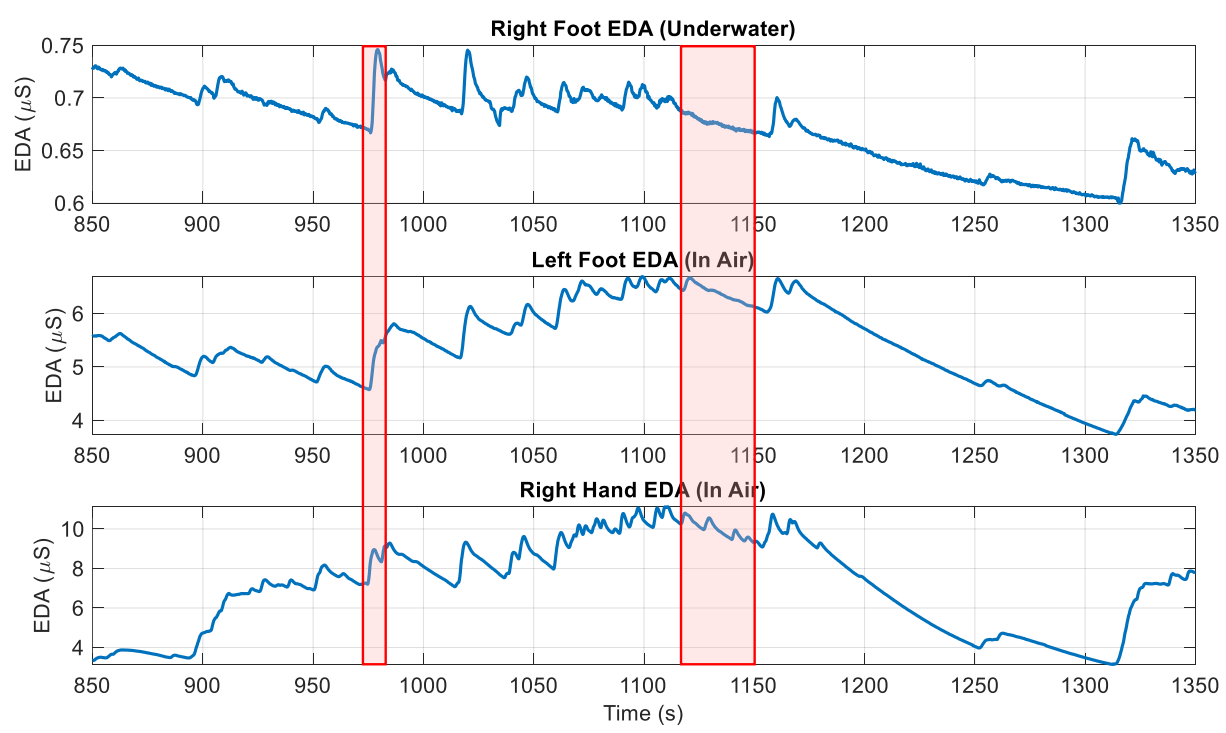

Fig. 1. Comparison of EDA signals across different measurement locations under underwater (right foot) and in-air (left foot and right hand) conditions. The red panels highlight time intervals (970–980s and 1120–1150s) where discrepancies in SCR patterns are evident. These observations highlight the inherent variability of EDA signals across different measurement sites and environmental conditions.

*B. Overall Framework*

Fig. 2 shows the overall architecture of the proposed EDA denoising framework, which is designed to effectively remove motion and environmental artifacts while enhancing downstream autonomic detection in real-world scenarios. The pipeline consists of three stages.

First, raw EDA signals are collected from divers in underwater environments, where they are corrupted by severe MAs and environmental noise. Second, a deployable 1D U-Net-style autoencoder is applied to denoise the corrupted EDA signals and reconstruct physiologically meaningful EDA components. Finally, the denoised EDA signals are used for the downstream task of predicting CNS-OT onset in divers.

## III. Dataset

For this study, we used four different types of datasets to train and evaluate the proposed denoising framework across diverse environmental and physiological conditions. Table I summarizes the key characteristics of each dataset, including the stimulation protocols, environmental conditions, and measurement locations. All EDA signals were collected at 100 Hz, except for Shimmer3 devices, which exhibit variable sampling rates up to 100 Hz, and subsequently downsampled to 4 Hz. Input segments were constructed using 512-sample windows (approximately 2 minutes).

Because EDA signals show asymmetric patterns across the measurement location [20], as shown in Fig. 1, simultaneous dual-site recordings cannot provide perfectly aligned input-target pairs for supervised training. Therefore, we exclusively used clean segments as training targets. To simulate realistic noisy conditions, these clean segments were augmented with either pure MA noise (from CMAD I) or synthetic artifacts (detailed in Section IV-A). For all training datasets (Public, CMAD II, and UMAC), an 80/20 subject-wise split was applied for training and validation, respectively.

*A. Public Datasets*

To train a generalized EDA denoising model under diverse physiological conditions, we compiled data from multiple established autonomic-inducing protocols [24]–[32], consistent with prior EDA denoising literature [16]. This combined dataset comprises 385 subjects, yielding 6,584 total segments (5,181 for training, 1,403 for validation).

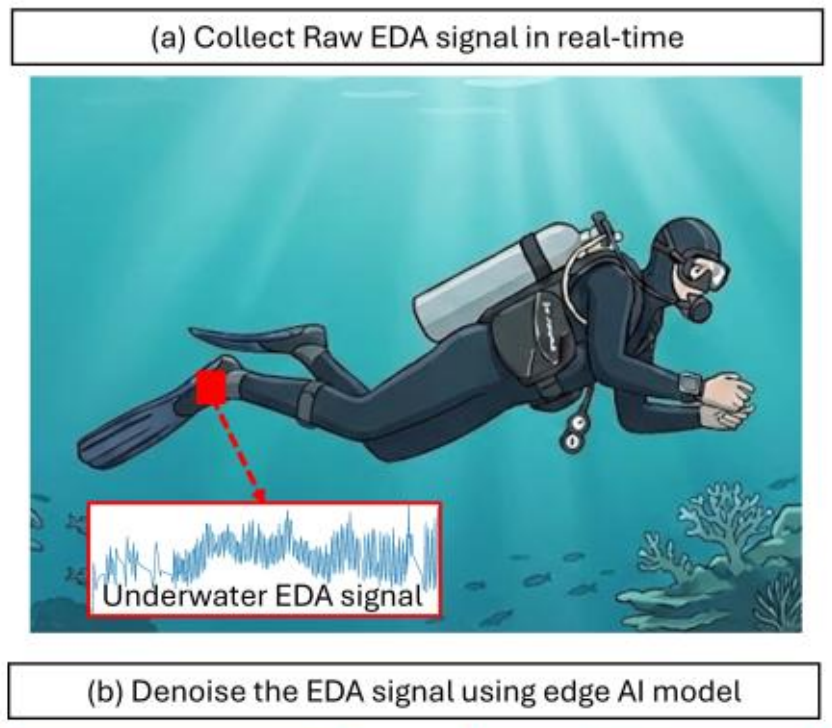

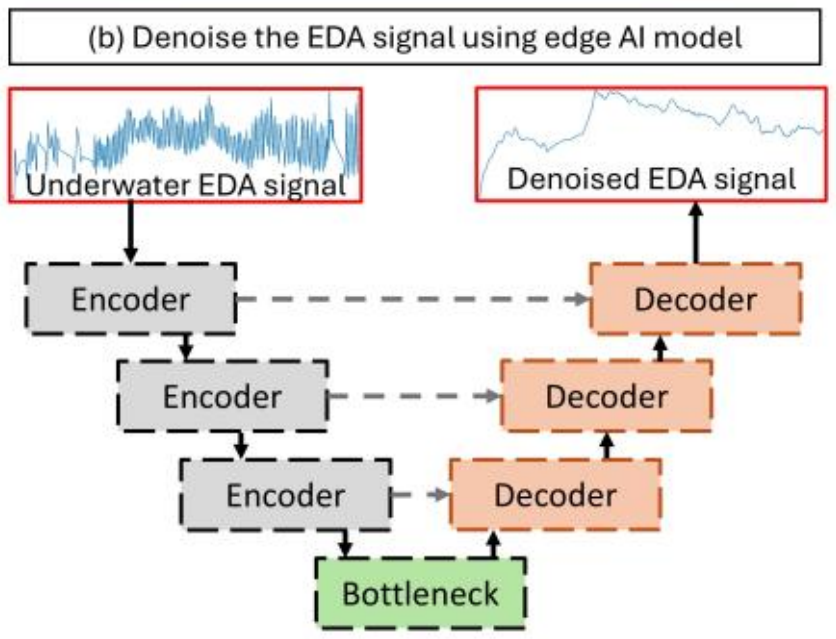

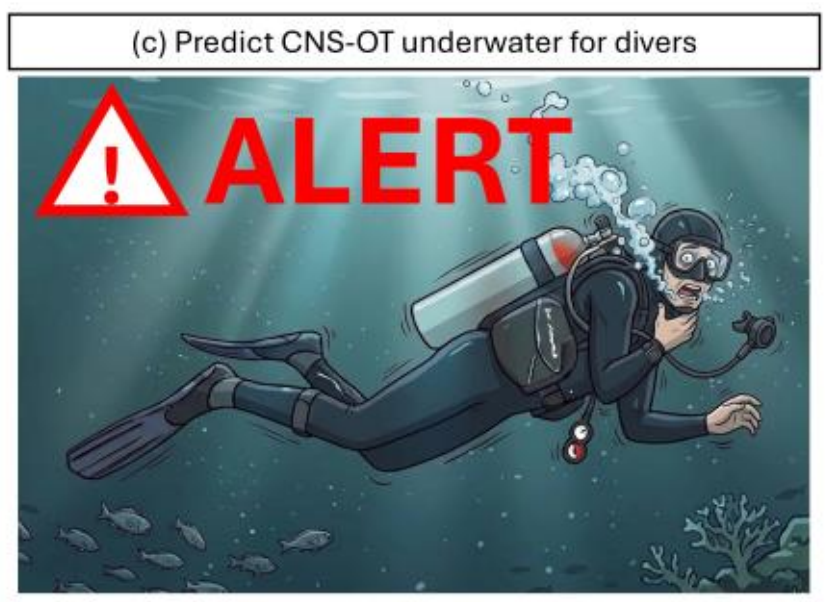

Fig. 2. Our proposed EDA denoising framework to effectively remove motion and noise artifacts while enhancing downstream autonomic detection in real-world scenarios.
*(Conceptual illustrations in Fig. 2 (a) and (c) were generated using Gemini 3 Flash Image based on text prompts for a scuba diver.)*

*B. Chonlab Motion Artifact Dataset (CMAD)*

CMAD I provides real-world leg motion artifacts recorded across a fixed resistor to remove physiological EDA components, yielding 60 pure MA segments used exclusively for noise augmentation [16]. CMAD II consists of simultaneous in-air EDA recordings from the immobile (left) and motion artifact–induced (right) hands of seven subjects. To avoid physiological asymmetry, only the 74 clean segments from the immobile hand were used for training and validation (58 for training, 16 for validation).

TABLE I
Summary of Datasets Used for Cross-Location and Harsh-Environment EDA Denoising

| Category | Public Dataset | CMAD I | CMAD II | UMAC Dataset | CNS-OT Dataset |
|---|---|---|---|---|---|
| Environment | In air | In air | In air | Underwater | Partial immersion (finger in air) |
| Measurement location | Finger | Leg (only motion dynamic) | Finger | Foot (left and right foot) | Finger |
| Sampling Rate | 100 Hz | 100 Hz | ~ 100 Hz | ~ 100 Hz | 100 Hz |
| Recording Duration | 21.3 ± 13.8 min | 1 hour | 15 min | 36 min | ~ 2 hours |
| Signal Condition | Clean | Motion Artifact | Motion Artifact | Underwater Motion Artifact + Cognitive Task | Underwater Motion Artifact + CNS-OT |
| Usage for Model Development | Training & Validation | Data Augmentation | Training & Validation | Training, Validation, & Testing | Independent Testing |
| Number of Subjects | 385 | 2 | 7 | 13 | 30 subjects (45 recordings) |
| Total Segments | 6,584 | 60 | 74 | 532 | 945 |
| Device | CED Micro1401 | Shimmer3 GSR+ Unit | Shimmer3 GSR+ Unit | Shimmer3 GSR+ Unit | ADInstruments PowerLab + GSR Amp FE116 |

### C. Underwater Motion Artifact Correction (UMAC) Dataset

To address environmental and location variability, the UMAC dataset includes continuous foot-based EDA from 13 subjects performing diving motions and cognitive tasks in an underwater environment. To collect the EDA signal underwater, a superhydrophobic-coated EDA electrode was used [33]. The 532 clean baseline segments from the immobile left foot were used for training and validation (404 for training, 128 for validation), while the 532 MA-corrupted segments from the active right foot were reserved for independent testing. All experimental procedures involving human subjects were approved by the Institutional Review Board (Protocol H23-0425).

### D. Central Nervous System Oxygen Toxicity (CNS-OT) Dataset

To evaluate the downstream clinical utility of the proposed framework, we used continuous EDA recordings from 30 divers exposed to hyperbaric environments (100% $O_2$ at 35 fsw) until CNS-OT symptoms appeared. This cohort comprises two subsets: an existing dataset of 12 subjects (18 recordings) previously detailed in [9], and a newly collected dataset of 18 subjects (27 recordings) acquired for this study.

To generate the evaluation segments for the downstream classification task, the continuous EDA signals were divided into two phases for each subject. The first 5 minutes of the experiment served as the baseline (non-CNS-OT) phase, yielding 6 segments per subject using a sliding window approach (512-sample window with 75% overlap). Conversely, the final 10 minutes before CNS-OT onset (first CNS-OT symptoms identified by a team of clinicians) served as the event (CNS-OT) prediction window, yielding 15 segments per subject. In total, 945 segments (270 baseline segments and 675 event segments) were extracted from the 45 independent recordings across 30 subjects. This entire dataset was strictly used for independent testing to provide an unbiased evaluation of the model's clinical generalizability. All new experimental procedures involving human subjects were approved by the Institutional Review Board (Protocol H25-0265).

## IV. Method

### A. Data Preprocessing and Augmentation

For the validation set, augmentation was performed before model training using all extracted EDA segments (512-sample windows) to generate noisy–target pairs. In contrast, for the training set, augmentation was applied dynamically during training. As detailed in Section III, physiological asymmetry prevents the use of simultaneous dual-site recordings for supervised training. Therefore, we exclusively used clean segments as the physiological ground truth (target) and generated corresponding noisy inputs via targeted data augmentation. Rather than applying simple, arbitrary noise such as Gaussian white noise, our augmentation strategy was designed to mimic the empirical characteristics of the real-world MAs observed in the CMAD and UMAC datasets. Table II summarizes the augmentation strategies.

TABLE II
SUMMARY OF AUGMENTATION STRATEGIES AND PARAMETERS

| Category | Augmentation Type | Parameter (Range) |
|---|---|---|
| Noise Artifacts | Real-world Extracted Motion Artifacts | Amplitude Scaling (0.7–1.3 × std) |
| | | Random Circular Shift |
| | Colored Noise | SNR (20–30 dB) |
| | | β (0–2) |
| Signal Distortion | Amplitude Clipping | Clipping (5 ~ 20%, upper saturation) |
| | Impulse Noise | Amplitude Scaling (5–30 × std)<br>Count (1–3 spikes)<br>Width (1–11 samples) |
| | Signal Shearing | Range (min–max)<br>Count (1–3 times) |

As detailed in Section III-B, real-world motion artifacts were extracted from the CMAD I, where physiological EDA components were removed using a fixed resistor setup. To prevent the model from memorizing specific noise templates, we applied random circular shifting to the extracted MA segments. Additionally, the amplitude of the added noise was dynamically scaled between 0.7 and 1.3, based on the standard deviation of the corresponding clean target segment.

Although amplitude scaling and circular shifting increased data variance, the limited pool of real-world MAs (60 segments) was insufficient to ensure robust model generalization. Therefore, we added synthetic colored noise (White, Pink, and Brown) to induce the model to learn the diverse, predominantly low-frequency characteristics of MAs. The intensity of this synthetic noise was empirically determined through validation. Specifically, applying a signal-to-noise ratio (SNR) between 20 and 30 dB during training yielded the best denoising performance (lowest root mean square error) when evaluated against a validation set of clean segments corrupted by real-world extracted MAs (CMAD I). This SNR range reflects training augmentation and does not represent real-world EDA SNR, which can be much lower (e.g., 0 dB or below).

In addition to noise artifacts, non-linear signal distortions, such as amplitude clipping, impulse noise, and signal shearing, were incorporated to model physical sensor or wire faults. These artifacts were selected by observations of real-world noisy segments. The parameters (the intensity and frequency of these distortions) were determined empirically.

To generate the final corrupted noisy inputs, we combined one noise artifact category with one signal distortion category (vide Table II). Fig. 3(a) shows representative examples of the mixed-noise input, clean target, and real MA-corrupted signal collected simultaneously (the latter two signals). By augmenting noise onto the clean target, the resulting signal mimics the noisy characteristics of real MA-corrupted EDA while preserving the underlying baseline characteristics of the clean signal, as shown in Fig. 3(a). In addition, Fig. 3(b) demonstrates that the augmented signals exhibit similar spectral characteristics to real MA-corrupted EDA. The training batches included 20% clean segments, ensuring the model learns to preserve the uncorrupted EDA signal, and 80% mixed-noise augmented segments. Conversely, to evaluate denoising capability, the validation set consisted entirely of mixed-noise signals.

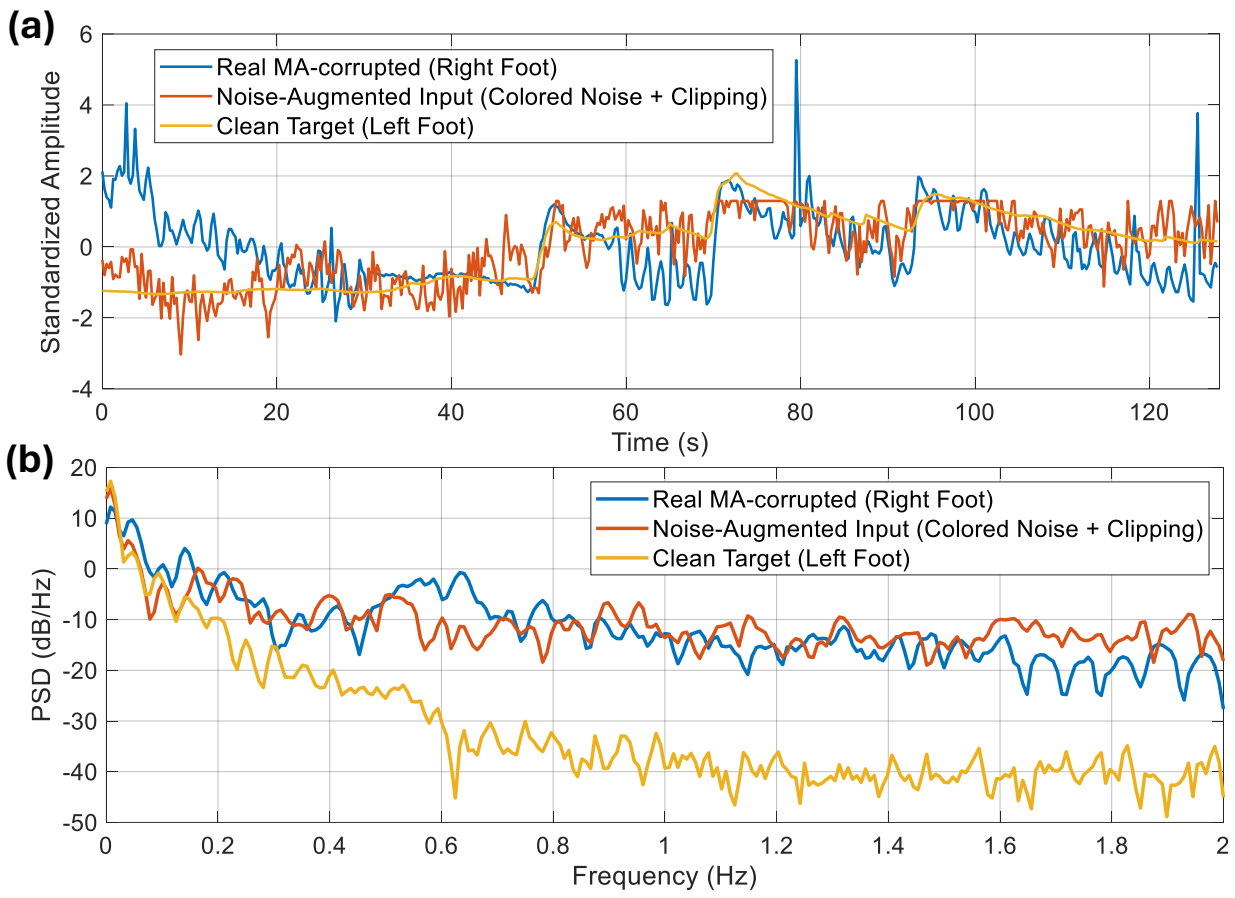

Fig. 3. Representative examples of the noise-augmented signal, clean target, and real MA-corrupted EDA collected simultaneously. (a) Time-domain comparison, demonstrating that the noise-augmented signal mimics real MA-corrupted EDA while preserving the underlying baseline dynamics of the clean EDA, in contrast to the asymmetric real MA-corrupted signal. (b) Power spectral density comparison showing that the augmented EDA exhibits similar spectral characteristics to real motion artifacts.

To prevent overfitting, dynamic augmentation was used during training, meaning the specific combination and parameters of the added noise were randomized at each epoch. For the validation set, we used static augmentation, where the noisy–target pairs were fixed prior to training to ensure a consistent evaluation benchmark. Finally, to stabilize the learning process and prevent gradient explosion, standardization (Z-score normalization) was applied to both the input and target segments. During testing and evaluation, the model's denoised outputs were inversely transformed (de-normalized) back to their original amplitude scales to allow for accurate physiological comparison and metric calculation.

### *B. Teacher Model for EDA Denoising*

Fig. 4(a) shows the proposed Teacher EDA denoising model. We adopt a hybrid CNN–Transformer architecture inspired by TransUNet [34], which combines a CNN-based encoder with a Transformer bottleneck.

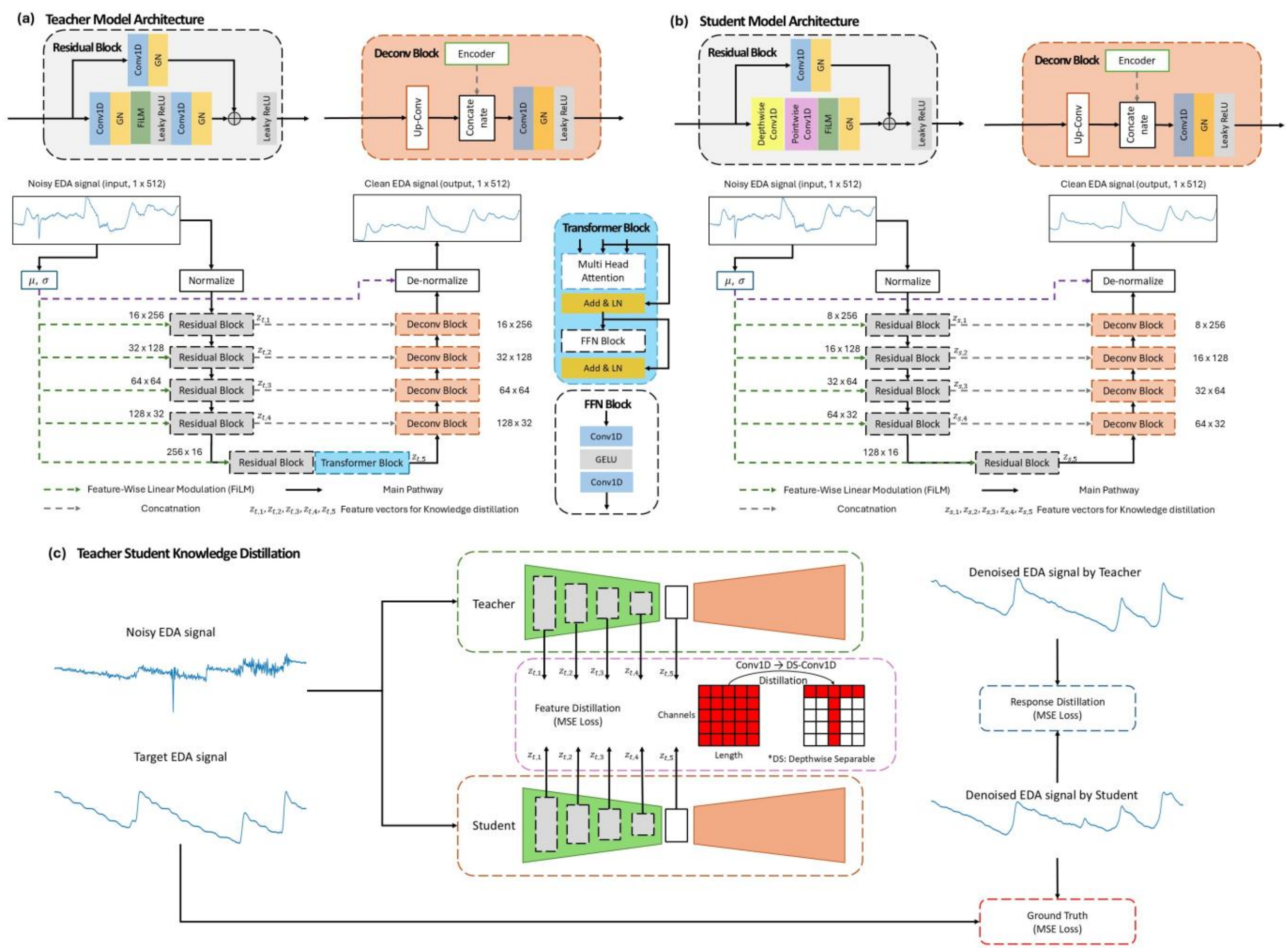

Fig. 4. Overview of the proposed EDA denoising framework. (a) Teacher model based on a hybrid CNN–Transformer architecture with FiLM-based feature modulation and a Transformer bottleneck for capturing long-range temporal dependencies. (b) Student model using depthwise separable convolutions (DS-Conv1D) for a lightweight model size in resource-constrained environments. (c) Knowledge distillation framework, where the student model is trained using both ground-truth reconstruction loss and distillation losses, including response and multi-level feature distillation from the teacher model.

As shown in Fig. 4(a), the noisy input $x \in \mathbb{R}^{1x512}$ first standardized (Z-score normalization) using the mean ($\mu$) and standard deviation ($\sigma$). The normalized signal is passed into the encoder, which consists of sequential residual blocks. Each residual block contains 1D convolutions (Conv1D) paired with Group Normalization (GN) and Leaky ReLU activation functions. To ensure the network dynamically adapts to the varying amplitude scales of different EDA placements (e.g., foot vs. hand) and environment (e.g., underwater vs. in air), a FiLM layer is integrated into the residual blocks [23].

The FiLM layer linearly modulates the feature maps within each residual block ( $x_i$ ), after the initial convolution and normalization. The scale ($\gamma_i$) and shift ($\beta_i$) factors for $i$-th block (where $i \in \{1, 2, 3, 4, 5\}$) are derived from the mean ($\mu$) and standard deviation ($\sigma$) of the input EDA segment. $\mu$ and $\sigma$ are passed through independent fully connected (FC) layers to compute the parameters:

$$\gamma_i = W_{\gamma_i}\mu + b_{\gamma_i} + 1 \quad (1)$$

$$\beta_i = W_{\beta_i}\sigma + b_{\beta_i} \quad (2)$$

where $W$ and $b$ represent the learned weights and biases of the FC layers. To stabilize early training, the weights and biases are initialized to zero, and a $+1$ offset is added to $\gamma_i$. Once derived, these parameters linearly modulate the intermediate feature map $x_i$:

$$x_i' = \gamma_i \cdot x_i + \beta_i \quad (3)$$

where $x_i'$ is the final output. This design allows the network to stably amplify or suppress features based on the physiological amplitude of the current segment.

The extracted features from the final encoder layer are then processed by a bottleneck Transformer Block. In this bottleneck, the sequence first passes through a multi-head attention module followed by layer normalization (LN) and a residual connection. Subsequently, the features are fed into a 1D convolutional feed-forward network (ConvFFN), which has been shown to better capture temporal dynamics in 1D physiological signals [35]. This is followed by GELU activation, a residual connection, and LN.

Finally, the decoder mirrors the encoder, utilizing transposed convolutions for upsampling and concatenating skip connections from the encoder to reconstruct the denoised signal. The final output is then de-normalized back to the original amplitude scale. This teacher model contains approximately 2.06M parameters (7.87 MB).

### C. Student Model for Knowledge Distillation

Fig. 4(b) shows the proposed lightweight student model for EDA signal denoising. To achieve a low computational time and memory usage deployable for wearable devices, the student architecture is motivated by MobileNet [36]. We replace the standard 1D convolutions in the encoder with 1D depth-wise separable convolutions (DS-Conv1D), comprising a depth-wise Conv1D followed by a pointwise Conv1D. Furthermore, the Transformer bottleneck is removed for computational efficiency. Finally, the channels across all layers are halved relative to the teacher model. Overall, these modifications significantly reduce the model parameters from 2.06M (7.87 MB) to 0.135M (0.51 MB).

To effectively mimic the teacher model's representations despite reduced parameters, a knowledge distillation strategy was employed [22]. The total training loss ($\mathcal{L}_{total}$) balances the standard reconstruction error against the clean target ($y$), which is first standardized ($y_{norm}$) using $\mu$ and $\sigma$ from noisy input, with distillation loss from the teacher:

$$\mathcal{L}_{total} = 0.5\mathcal{L}_{MSE}(y_{norm}, \hat{y}_s) + 0.5\mathcal{L}_{KD} \quad (4)$$

where $\hat{y}_s$ is the denoised output of the student model. The knowledge distillation loss ($\mathcal{L}_{KD}$) is defined as a combination of response distillation and feature distillation:

$$\mathcal{L}_{KD} = \mathcal{L}_{response} + 0.3\mathcal{L}_{feature} \quad (5)$$

The response distillation ($\mathcal{L}_{response}$) minimizes the mean squared error (MSE) between the denoised output of the teacher model ($\hat{y}_t$) and student model ($\hat{y}_s$):

$$\mathcal{L}_{response} = MSE(\hat{y}_t, \hat{y}_s) \quad (6)$$

Simultaneously, the feature distillation ($\mathcal{L}_{feature}$) ensures the student model learns the complex representation of the teacher's deep layers. Both feature maps in the student and teacher models ($z_{t,i}, z_{s,i}$ ) are extracted from each of $i$-th encoder blocks (where $i \in \{1, 2, 3, 4, 5\}$ ). Given that the channel dimensions between the teacher and student differ, a 1x1 convolutional projection layer ($\phi_i$) is applied to linearly map the student's feature maps ($z_{s,i}$) to teacher's dimensions ($z_{t,i}$). The feature distillation loss is calculated as the sum of the MSE across five encoder blocks:

$$\mathcal{L}_{feature} = \sum_{i=1}^{5} MSE(z_{t,i}, \phi_i(z_{s,i})) \quad (7)$$

By optimizing both the ground-truth reconstruction ($\mathcal{L}_{response}$) and the multi-level distillation ($\mathcal{L}_{KD}$), the student model generates a high-fidelity denoised signal ( $\hat{y}_s$ ) while maintaining low memory usage and computational efficiency, making it suitable for deployment on resource-constrained wearable devices.

### D. CNS-OT Prediction Framework

An example use case of the proposed denoising framework is to facilitate the real-time, early prediction of CNS-OT in operational divers. First, raw EDA segments are evaluated by the MA detector (in Section IV-B). Segments classified as unrecoverable noise are discarded to prevent false alarms, while recoverable segments are processed by the proposed KD student denoising model. The denoised EDA signals are subsequently fed into the downstream prediction algorithm.

CNS-OT risk is evaluated by computing a risk score derived from phasic drivers, which represent discrete sudomotor nerve impulses underlying SCRs, using methodologies established in prior works [36]–[37]. For model evaluation, the CNS-OT dataset was divided into two distinct physiological states: a baseline segment and a pre-CNS-OT segment. The baseline segment represents the diver's baseline physiology, extracted from the first 5 minutes of the experiment. Conversely, the pre-CNS-OT segment captures the critical physiological deterioration phase, extracted from the final 10 minutes preceding the clinical onset of CNS-OT symptoms. By monitoring the physiological changes within the denoised EDA stream, the proposed framework is designed to provide early warning of CNS-OT onset before the appearance of clinical symptoms.

### *E. Evaluation Metrics*

Signal reconstruction performance on the augmented dataset was evaluated against the clean target using mean absolute error (MAE), root mean square error (RMSE), Pearson correlation coefficient (PCC), and SNR improvement (SNR). To ensure accurate physiological comparison, all reconstruction metrics were computed on the de-normalized signal. Let $x$, $y$, and $\tilde{y}$ denote the raw noisy input, the raw clean target, and the de-normalized model output, respectively. The SNR improvement was calculated as:

$$SNR_{imp} = 10log_{10}\left(\frac{\sum_i(x[i]-y[i])^2}{\sum_i(\tilde{y}[i]-y[i])^2}\right) \tag{8}$$

For the real-world UMAC dataset, reconstruction performance was evaluated by comparing the extracted phasic components (SCRs) from the clean left-foot signal with those from the noisy right-foot signal, using the same metrics (MAE, RMSE, PCC, and SNR improvement).

Clinical prediction performance was evaluated under a leave-one-subject-out (LOSO) cross-validation. First, the segment-level evaluation was conducted to determine decision thresholds and assess discriminative performance. The area under the receiver operating characteristic curve (AUROC) was computed and to strictly penalize false alarms in practical settings, the segment-level decision thresholds (phasic driver) were selected from the training fold to achieve 90% specificity and subsequently applied to the held-out subject. Building on this, recording-level evaluation was performed to reflect practical real-world scenarios, where detecting at least one event is critical for early warning of CNS-OT onset. Sensitivity, specificity, and accuracy were computed at the recording level. Subject-level sensitivity (early prediction rate) was defined as the proportion of these recordings with at least one correct prediction during the 15-segment CNS-OT window. Specificity was defined as the proportion of recordings with zero false positive alarms during the 6-segment baseline period. For subjects with multiple recordings, these recording-level metrics were averaged to yield a single representative value per subject. The final AUROC, sensitivity, specificity, and accuracy are reported as macro-averages across all subjects. Finally, early prediction time is reported as the median of the subject-level average lead times.

To compare downstream CNS-OT prediction performance across denoising methods, statistical analyses were performed on subject-wise LOSO metrics using linear mixed-effects models (LMER), with denoising method as a fixed effect and subject as a random effect. Post hoc pairwise comparisons were conducted using estimated marginal means with Tukey adjustment for multiple comparisons.

To assess deployment feasibility for wearable IoT systems, computational efficiency was evaluated in terms of total parameters, model size, floating-point operations (FLOPs), and inference latency.

### *F. Training Strategy*

The proposed model was trained using the AdamW optimizer with a weight decay of 1e-4 and a batch size of 16. The initial learning rate was 1e-3 and gradually decreased to 1e-6 using a cosine annealing scheduler over 200 epochs. The teacher model was trained using an MSE loss to optimize signal reconstruction. The student model was trained using a combination of reconstruction and KD losses as described in Section IV-D. All models were trained 200 epochs under the same optimization settings to ensure a fair comparison. The best-performing model was selected based on the lowest MAE on the validation set. All experiments were implemented in PyTorch and trained on an NVIDIA A100 GPU using Google Colab.

## V. RESULTS

### *A. Ablation Study Results*

To validate the architectural design of the proposed teacher model, an ablation study was conducted by comparing a baseline CNN-based U-Net (without FiLM and Transformer block in Fig. 4(a)) with variants incorporating the FiLM layer and the bottleneck Transformer block, as summarized in Table III. These results were obtained using the augmented noisy training/validation dataset described in Table II, which was constructed from clean EDA signals from the public dataset, CMAD II, and UMAC dataset combined with synthetic noise and real-world MAs (CMAD I).

The baseline U-Net yields a moderate denoising performance with an MAE of 0.144 and an SNR of 12.27. FiLM layer (+ FiLM) provides a marginal improvement in reconstruction accuracy, reducing the MAE to 0.141 and the RMSE from 0.182 to 0.179. Notably, the FiLM layer reduces the standard deviations across these error metrics (e.g., MAE standard deviation drops from 0.277 to 0.222), indicating that feature conditioning leads to more stable and consistent denoising across varying signal conditions.

TABLE III
ABLATION STUDY RESULTS OF THE TEACHER MODEL

| Model | MAE | RMSE | PCC | SNR |
|---|---|---|---|---|
| Baseline U-Net | 0.144 ± 0.277 | 0.182 ± 0.319 | 0.896 ± 0.173 | 12.27 ± 6.443 |
| + FiLM | 0.141 ± 0.220 | 0.179 ± 0.266 | 0.893 ± 0.179 | 12.28 ± 6.357 |
| + Transformer Block | 0.134 ± 0.268 | 0.169 ± 0.304 | 0.906 ± 0.166 | 12.93 ± 6.546 |
| **Proposed Model** | **0.125 ± 0.187** | **0.160 ± 0.227** | **0.909 ± 0.169** | **13.17 ± 6.744** |

The integration of the Transformer block (+ Transformer Block) improves denoising performance by capturing the overall signal morphology, as reflected in the improved metrics: MAE decreases from 0.144 to 0.134, RMSE from 0.182 to 0.169, PCC increases from 0.896 to 0.906, and SNR improves from 12.27 to 13.17 dB. These results demonstrate the Transformer's ability to effectively model long-range temporal dependencies in EDA signals.

Finally, the teacher model (Proposed Model), which combines both the FiLM layer and the Transformer Block, achieves the best overall performance across all evaluated metrics. It reaches the lowest error rates (MAE of 0.125, RMSE of 0.160) and the highest signal fidelity (PCC of 0.909, SNR of 13.17). These results demonstrate that the two modules (FiLM and Transformer Block) operate synergistically to enable robust, high-fidelity signal reconstruction, thereby validating the architectural design of the teacher model.

### *B. Knowledge Distillation Results*

To evaluate the efficacy of the proposed KD framework, we compared the denoising performance of four models: the teacher model, the student model without KD, the Student model with KD, and a prior baseline EDA denoising model, U-Net [16]. To ensure a fair comparison, the same training strategy and dataset were utilized across all models.

Fig. 5 presents an example of the denoising results on a signal corrupted by colored noise and signal shearing artifacts. The models exhibit varying levels of robustness to these distortions. For instance, at approximately the 120-second mark (second red highlighted zone), a severe signal shearing artifact corrupts the input signal. In this case, the student model without KD struggles to reconstruct the target signal, whereas the teacher model effectively suppresses the artifact. The student model with KD and U-Net [16] partially mitigate the shearing artifact; however, the U-Net [16] exhibits limitations in removing high-frequency noise components present in the EDA signal. At around the 30-second mark (first red highlighted zone), all models show minor deviations from the target signal when filtering colored noise. This is likely due to the overlap between the frequency band of the augmented colored noise and the intrinsic frequency components of SCRs, making complete separation inherently challenging.

Table IV summarizes the denoising performance of the evaluated models in terms of MAE, RMSE, PCC, and SNR, using the same augmented noisy training and validation dataset used in Table III. The U-Net [16] (2.55 MB) demonstrates moderate performance, achieving an MAE of 0.215 and an SNR of 9.198 dB. Notably, the student model without KD, despite its significantly smaller size (0.51 MB), outperforms U-Net across all metrics (MAE: 0.192 vs. 0.215, RMSE: 0.243 vs. 0.283, PCC: 0.840 vs. 0.815, and SNR: 9.828 vs. 9.198 dB), highlighting the efficiency of the proposed lightweight architecture. However, a noticeable performance gap remains between the student model without KD and the teacher model (7.87 MB), which achieves superior denoising performance.

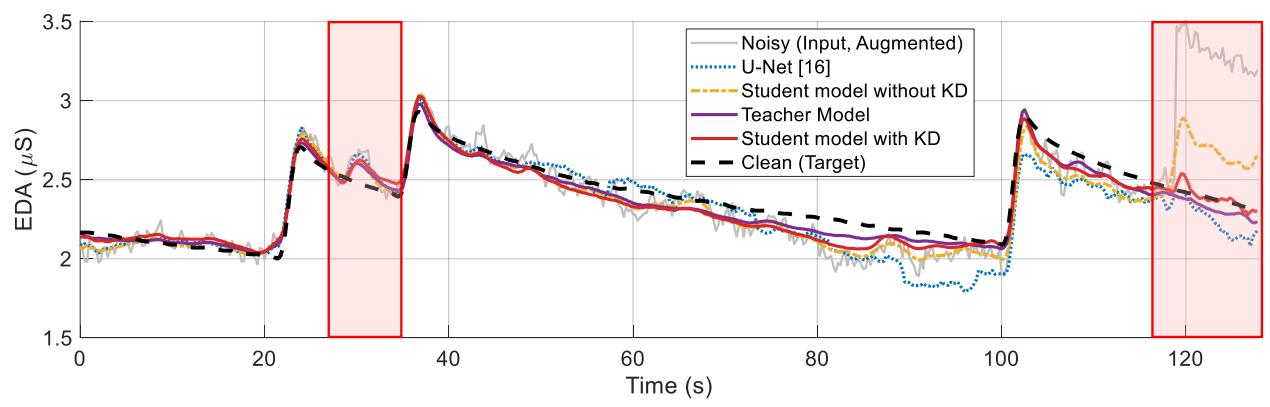

Fig. 5. An example comparison of EDA denoising results on a signal corrupted by colored noise and shearing artifacts. The red panels highlight specific periods of severe signal distortion. The student model with KD mimics the teacher model in suppressing severe shearing (e.g., around 120 s), outperforming the U-Net [16] and the student without KD.

TABLE IV
DENOISING PERFORMANCE COMPARISON OF THE STUDENT MODEL WITH AND WITHOUT KNOWLEDGE DISTILLATION

| Model | MAE | RMSE | PCC | SNR |
|---|---|---|---|---|
| U-Net [16] (2.55 MB) | 0.215 ± 0.384 | 0.283 ± 0.474 | 0.815 ± 0.268 | 9.198 ± 5.985 |
| Student Model without KD (0.51MB) | 0.192 ± 0.307 | 0.243 ± 0.355 | 0.840 ± 0.230 | 9.828 ± 5.936 |
| Student Model with KD (0.51MB) | 0.144 ± 0.240 | 0.185 ± 0.285 | 0.890 ± 0.182 | 12.08 ± 6.423 |
| **Teacher Model (7.87 MB)** | **0.125 ± 0.187** | **0.160 ± 0.227** | **0.909 ± 0.169** | **13.17 ± 6.744** |

By applying the proposed Teacher–Student KD strategy, the student model with KD effectively narrows this gap without increasing model complexity. Specifically, the KD-trained student reduces the MAE from 0.192 to 0.144 and the RMSE from 0.243 to 0.185, while improving the PCC from 0.840 to 0.890 and the SNR from 9.828 to 12.08 dB compared to the student model without KD. These results demonstrate that the KD framework successfully transfers the representational capacity of the teacher model to the lightweight student model, enabling robust EDA denoising suitable for memory-constrained edge deployment.

### *C. Underwater EDA Reconstruction Results*

To evaluate the signal reconstruction performance in highly corrupted real-world scenarios, we compared four different EDA MA removal methods on the UMAC dataset. The baselines include an exponential filtering method ($\alpha$ = 0.8) [39], a SWT-based method utilizing Laplace modeling [15], the prior U-Net baseline [16], and our proposed KD-based student model.

A key challenge in evaluating real-world wearable EDA signals is the absence of a reliable ground truth. Fig. 6(a) illustrates EDA signals simultaneously collected from two locations (left and right foot). Even when MAs are confined to the right foot underwater, the signals from the two sites exhibit different morphologies. For example, the right foot amplitude ranges from 11.2 to 12.2 μS, whereas the cleaner left foot signal ranges from 0.4 to 0.8 μS. Consequently, directly computing error metrics on raw EDA signals across different measurement sites is inappropriate. However, although the baseline (tonic) components vary between sites, the SCRs remain more temporally synchronized, as shown in Fig. 6 (b) and (c).

Since EDA signals are typically used after SCR extraction in downstream applications, we evaluate reconstruction performance based on the recovered phasic components. Specifically, we employ two established EDA decomposition (Decomp.) methods: cvxEDA [37], a widely used convex optimization–based approach, and ospEDA [38], an orthogonal subspace projection–based method designed for robustness under noisy conditions. Note that ospEDA outperformed existing decomposition methods in accurately estimating phasic drivers under real-world noise conditions [38].

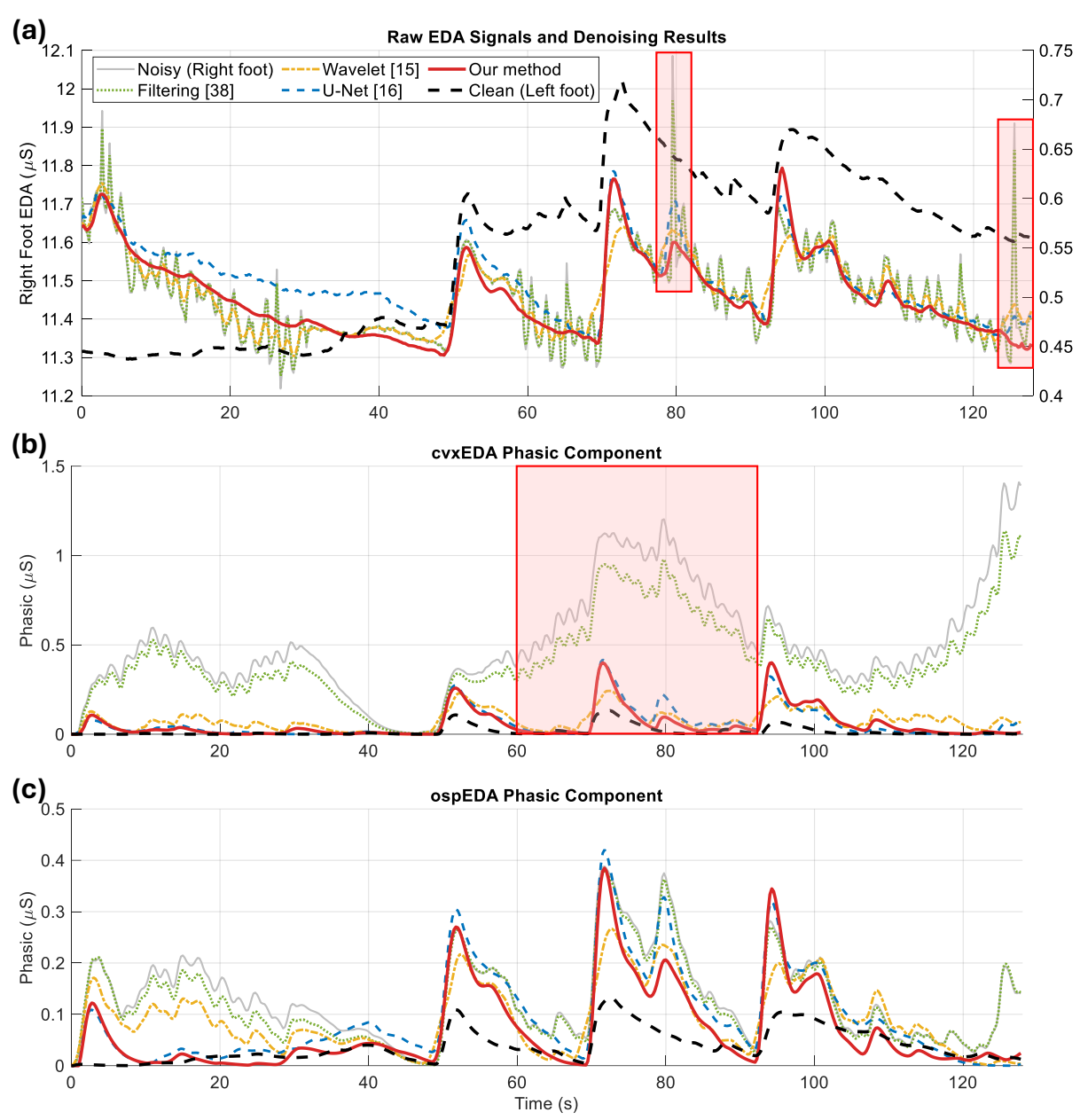

Fig. 6. Representative comparison of EDA reconstruction in an underwater scenario (UMAC dataset). (a) Raw EDA signals showing MA-corrupted right foot vs. clean left foot reference and denoised outputs. (b) Phasic components extracted using cvxEDA. (c) Phasic components extracted using ospEDA. The red panels highlight significant impulse noises and phasic deviation.

Fig. 6(b) and Fig. 6(c) compare the decomposed SCRs (phasic components) of the denoised signals (via various filtering and our proposed KD-student model) against the clean reference. As shown in Fig. 6(b) using cvxEDA, the noisy signal and the exponential filtering method [39] exhibit noticeable deviations from the clean reference. In contrast, the wavelet-based method [15], U-Net [16], and the proposed method (KD-student model) effectively suppress environmental noise, producing phasic components that more closely follow the clean left-foot reference. Fig. 6(c) shows a similar trend using ospEDA. Compared to cvxEDA, ospEDA further reduces deviations in the noisy and exponential filtering results, particularly in the 60–90 s interval (red rectangular zone in Fig. 6(b)). This improved suppression enables a clearer assessment of denoising performance. Notably, the proposed method demonstrates performance in mitigating impulse noise, as evidenced by the reduced amplitude of false peaks, particularly around 125 s, while partially attenuating artifacts near 80 s (red rectangular zones in Fig. 6(a).

Table V presents the SCR reconstruction performance on the UMAC dataset using two decomposition methods, cvxEDA and ospEDA. Under the cvxEDA framework, the raw EDA (without denoising) and exponential filtering [39] methods exhibit substantially degraded performance, with significantly higher MAE (2.809 and 1.798) and RMSE (3.294 and 2.140), along with lower PCC values (0.164 and 0.170), indicating their inability to effectively suppress MAs. In contrast, the proposed method (KD-student model) achieves the best overall performance, yielding the lowest MAE (0.215) and RMSE (0.298), along with improved PCC (0.258) and the highest SNR (13.04 dB), outperforming wavelet-based denoising [15] and U-Net [16].

TABLE V
SCR Reconstruction performance on the UMAC dataset

| Decomp. | Method | MAE | RMSE | PCC | SNR |
|---|---|---|---|---|---|
| cvxEDA | Without denoising | 2.809 ± 22.24 | 3.294 ± 25.24 | 0.164 ± 0.404 | - |
| | **Our model** | **0.215 ± 3.464** | **0.298 ± 4.173** | **0.258 ± 0.415** | **13.04 ± 22.35** |
| | Filtering [39] | 1.798 ± 13.69 | 2.140 ± 15.46 | 0.170 ± 0.404 | 0.634 ± 2.012 |
| | Wavelet [15] | 0.261 ± 3.491 | 0.351 ± 4.199 | 0.212 ± 0.399 | 7.134 ± 12.30 |
| | U-Net [16] | 0.237 ± 3.466 | 0.325 ± 4.177 | 0.228 ± 0.404 | 9.176 ± 15.93 |
| ospEDA | Without denoising | 0.179 ± 0.654 | 0.223 ± 0.773 | **0.224 ± 0.358** | - |
| | **Our model** | **0.065 ± 0.139** | **0.083 ± 0.171** | 0.219 ± 0.369 | **5.582 ± 7.464** |
| | Filtering [39] | 0.175 ± 0.675 | 0.219 ± 0.793 | 0.214 ± 0.362 | 0.386 ± 1.345 |
| | Wavelet [15] | 0.103 ± 0.285 | 0.132 ± 0.369 | 0.185 ± 0.362 | 2.401 ± 4.972 |
| | U-Net [16] | 0.073 ± 0.172 | 0.092 ± 0.209 | 0.195 ± 0.372 | 4.759 ± 7.029 |

Under the ospEDA framework, which is more robust to noise (compared to cvxEDA), most methods show improved performance in terms of MAE and RMSE, including the raw EDA (without denoising) case. Nevertheless, the proposed method consistently achieves the best results across all metrics, with the lowest MAE (0.065) and RMSE (0.083), as well as the highest PCC (0.219) and SNR (5.582 dB). These findings indicate that while ospEDA mitigates noise at the decomposition stage, the proposed method further enhances reconstruction quality.

### D. CNS-OT Prediction Results

To evaluate the downstream clinical applicability of the proposed denoising framework, we assessed its performance on CNS-OT prediction. For each subject, the first 5 minutes were used as the baseline (non–CNS-OT) segment, while the final 10 minutes were defined as the event window for CNS-OT prediction. Classification was performed using the peak amplitude of the phasic drivers (threshold) derived from SCRs.

Table VI presents the CNS-OT prediction results across different denoising methods and decomposition frameworks. Under both cvxEDA and ospEDA, the proposed method achieved the strong and consistent performance across multiple evaluation metrics.

Under the cvxEDA framework, the proposed method achieved the highest AUROC (0.806), indicating improved segment-level discriminative capability, with statistical significance compared to the without denoising baseline ($p < 0.05$). In addition, the proposed model significantly improved sensitivity (0.700) and accuracy (0.775) ($p < 0.05$), demonstrating enhanced CNS-OT prediction performance at both the segment and subject levels. Notably, the early prediction time was extended from 4.33 minutes to 6.93

TABLE VI

CNS-OT PREDICTION RESULTS ACROSS DIFFERENT DENOISING METHODS AND DECOMPOSITION FRAMEWORKS UNDER LOSO CROSS-VALIDATION.

| Decomp. | Method | Sensitivity | Specificity | Accuracy | AUROC | Early Prediction Time | Threshold |
|---|---|---|---|---|---|---|---|
| cvxEDA | Without denoising | 0.433 ± 0.450 | 0.767 ± 0.410 | 0.600 ± 0.283 | 0.671 ± 0.290 | 4.33 (1.87 – 7.20) min | 34.14 ± 1.203 |
| | **Our model** | 0.700 ± 0.428* | **0.850 ± 0.298** | **0.775 ± 0.265*** | **0.806 ± 0.240*** | **6.93 (1.60 – 7.47) min** | 4.049 ± 0.198 |
| | Filtering [39] | 0.467 ± 0.454 | 0.800 ± 0.385 | 0.633 ± 0.269 | 0.652 ± 0.295 | 3.73 (0.53 – 7.07) min | 31.11 ± 0.864 |
| | Wavelet [15] | 0.633 ± 0.434 | 0.833 ± 0.330 | 0.733 ± 0.245 | 0.771 ± 0.285 | **6.93 (3.20 – 7.47) min** | 3.786 ± 0.145 |
| | U-Net [16] | **0.717 ± 0.409*** | 0.833 ± 0.330 | **0.775 ± 0.249*** | 0.777 ± 0.256* | **6.93 (2.13 – 7.47) min** | 4.974 ± 0.296 |
| ospEDA | Without denoising | 0.550 ± 0.461 | 0.833 ± 0.356 | 0.692 ± 0.252 | 0.718 ± 0.275 | 5.07 (0.80 – 7.47) min | 2.263 ± 0.114 |
| | **Our model** | **0.767 ± 0.388*** | 0.867 ± 0.292 | **0.817 ± 0.207*** | **0.794 ± 0.247** | **6.93 (1.07 – 7.47) min** | 0.721 ± 0.031 |
| | Filtering [39] | 0.550 ± 0.461 | 0.800 ± 0.385 | 0.675 ± 0.280 | 0.721 ± 0.275 | 5.07 (0.80 – 7.47) min | 2.141 ± 0.111 |
| | Wavelet [15] | 0.617 ± 0.449 | 0.850 ± 0.326 | 0.733 ± 0.254 | 0.778 ± 0.276 | **6.93 (3.60 – 7.47) min** | 0.798 ± 0.078 |
| | U-Net [16] | 0.733 ± 0.430 | **0.883 ± 0.284** | 0.808 ± 0.224 | 0.787 ± 0.255 | **6.93 (1.60 – 7.47) min** | 0.672 ± 0.046 |

Note: Values are reported as mean ± SD across subjects under LOSO cross-validation, except Pred Time, which is shown as median (Q1–Q3) among successful predictions only. For each decomposition method, asterisks denote statistical significance compared with the baseline (without denoising): $p < 0.05$ (*).

minutes.

Similarly, under the ospEDA framework, the proposed method maintained strong performance, achieving the highest sensitivity (0.767) and accuracy (0.817), both statistically significant compared to the without denoising baseline ($p < 0.05$), while also demonstrating a competitive AUROC (0.794). The early prediction time remained consistently high (6.93 min), further supporting the robustness of the proposed denoising method across different decomposition frameworks.

Across both frameworks, the baseline (without denoising) shows substantially lower sensitivity and AUROC, indicating that MAs and environmental noise significantly degrade downstream prediction performance. Across all decomposition methods, a similar performance trend was observed: the proposed model achieved the best performance, followed by U-Net, wavelet-based methods, filtering, and finally the without-denoising baseline. This trend further confirms that effective denoising plays a critical role in improving downstream CNS-OT prediction.

Fig. 7 provides a visual comparison between raw and denoised EDA signals from the CNS-OT dataset. The red rectangular zones indicate segments corrupted by MAs, which are effectively suppressed by the proposed model. The denoised signal preserves the underlying SCRs while removing abrupt noise fluctuations (e.g., impulsive artifacts) present in the raw input. Notably, artifact intensity tends to increase as subjects approach CNS-OT onset due to discomfort and movement. The proposed method mitigates these effects, enabling more stable extraction of phasic drivers and contributing to more reliable downstream CNS-OT prediction.

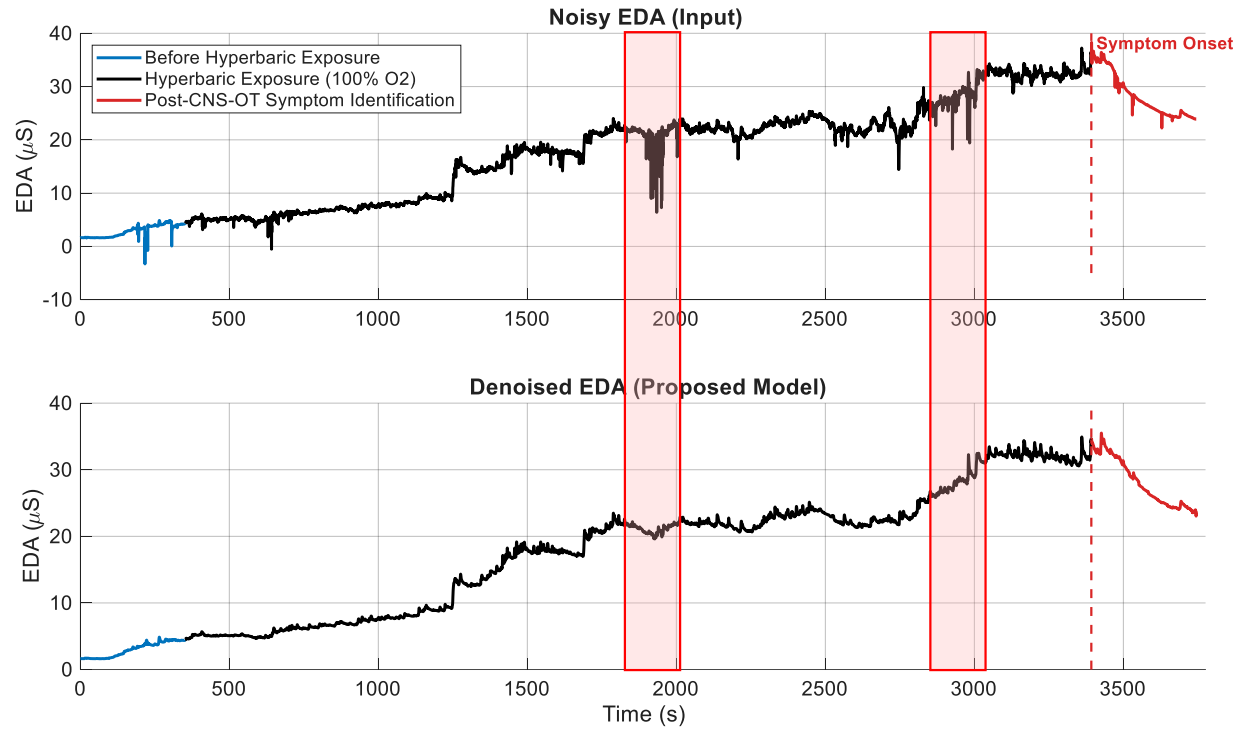

Fig. 7. Comparison of raw and denoised EDA signals from the CNS-OT dataset. The upper panel shows the noisy input EDA, while the lower panel shows the denoised signal obtained using the proposed student model. Red panels highlight segments affected by MAs, which are effectively suppressed in the denoised signal. The dashed vertical line indicates the onset of CNS-OT symptoms.

In conclusion, these results demonstrate that severe environmental noise and MAs can significantly degrade downstream clinical prediction; however, the proposed lightweight model effectively recovers the critical phasic information (SCRs) required for accurate and reliable CNS-OT prediction.

### *E. Deployment Feasibility in Wearable IoT*

To assess practical deployment feasibility in wearable IoT systems, we evaluate the computational complexity and inference latency of the teacher, student, and U-Net [16] models. Table VII summarizes the parameter count, model size, FLOPs, and CPU inference latency for each model.

While the teacher model achieves the highest denoising performance, its substantial computational cost (2.063M parameters, 7.87 MB model size, and 105.1 MFLOPs) limits its suitability for resource-constrained edge devices. In contrast, the student model significantly reduces complexity to 0.135M parameters, 0.51 MB, and 11.61 MFLOPs, corresponding to approximately 15×, 15×, and 9× reductions in parameters, model size, and FLOPs, respectively. Although this reduction introduces some performance degradation (Table IV), incorporating KD enables the student model to recover comparable denoising performance without increasing its computational footprint.

To further assess real-time feasibility, inference latency was measured on a CPU environment. The U-Net [16] demonstrates a lower inference latency (1.81 ms) than the student model (2.94 ms); however, its large storage footprint (2.55 MB) presents a significant limitation for memory-constrained embedded platforms. As noted in prior tiny machine learning studies, deploying deep learning on microcontroller units (MCUs) is

TABLE VII

COMPLEXITY AND INFERENCE LATENCY COMPARISON

| Model | Parameters (M) | Model Size (MB) | FLOPs (M) | Inference Latency (ms) |
|---|---|---|---|---|
| Teacher Model | 2.063 | 7.87 | 105.1 | 5.30 |
| **Student Model** | **0.135** | **0.51** | **11.61** | 2.94 |
| U-Net [16] | 0.670 | 2.55 | 57.96 | **1.81** |

*Storage size is calculated using base-2 representation (1 MB = 1024 KB), representing the static flash memory footprint required for device deployment.

** Inference latency is averaged over 1,000 inference iterations with a batch size of 1, measured on a CPU environment (Google Colab), serving as a proxy for execution time.

primarily constrained by limited on-chip memory (SRAM) and storage (Flash), making model size a critical bottleneck [21].

With a compact size of 0.51 MB, the proposed student model fits within the typical 1 MB Flash capacity of widely used MCUs (e.g., STM32F412, STM32F746, and STM32F765) [21]. Moreover, its 2.94 ms inference latency remains sufficiently low for real-time physiological signal processing. Overall, the KD-based student framework achieves a favorable balance between denoising performance and computational efficiency, making it suitable for deployment in wearable IoT applications.

## VI. Discussion

This study presents a comprehensive evaluation framework for EDA denoising that extends beyond EDA signal reconstruction metrics by quantifying its impact on a real-world downstream clinical application, specifically CNS-OT prediction. Furthermore, the proposed framework is designed to generalize across multiple measurement sites (e.g., foot and hand) and extreme operational environments, thereby bridging the gap between controlled laboratory conditions (in air) and harsh real-world deployments (underwater).

A fundamental challenge in wearable EDA signal processing is the substantial variability in signal amplitude and morphology, arising from differences in measurement location, environmental conditions (e.g., temperature), and subject-dependent physiological factors [17]–[20]. These variabilities influence both tonic (baseline) and phasic (SCRs) components, making it difficult for denoising models to generalize across heterogeneous conditions. In this work, this challenge is addressed through a normalization and de-normalization strategy that preserves the baseline amplitude while allowing the model to focus on EDA morphology. In contrast, static standardization, such as applying scaling parameters ($\mu$ and $\sigma$) computed from the training data, may perform adequately under controlled conditions but can fail in low amplitude environments such as underwater settings, where SCRs are attenuated or suppressed.

To further address amplitude variability, the proposed model incorporates FiLM, which enables adaptive feature scaling conditioned on input signal statistics ( $\mu$ and $\sigma$ ). This mechanism allows the model to dynamically adjust to varying amplitude distributions across measurement sites and environments, improving robustness under both in-air and underwater conditions. These results suggest that separating amplitude normalization from morphology learning is important for a generalized EDA model across diverse real-world scenarios.

Another critical challenge is the accurate reconstruction of SCRs under real-world noise conditions. Conventional filtering and wavelet-based MA removal methods primarily focus on eliminating high-frequency noise and often fail to address noise components that overlap with the SCR frequency band. As shown in Fig. 6(a), low-frequency environmental noise remains largely unremoved by these approaches. In contrast, the proposed model is trained using a diverse data augmentation strategy that incorporates realistic artifacts, including clipping, impulse noise, and shearing, as well as distortions that mimic SCR dynamics, such as colored noise and real-world extracted motion artifacts. This approach enables the model to learn to distinguish true SCRs from noise, resulting in more robust reconstruction under harsh environmental conditions and severe MAs.

From a deployment perspective, wearable IoT systems impose strict constraints on memory and computational resources. While the teacher model achieves superior denoising performance, its high complexity limits its applicability in embedded systems. To address this, a KD framework is employed to transfer feature representations from the teacher model to a lightweight student model. The resulting student model achieves a compact size of 0.51 MB while maintaining competitive performance, making it suitable for deployment on resource-constrained MCUs with limited Flash memory (e.g., < 1 MB) [21].

Importantly, the proposed framework demonstrates that improved signal reconstruction can translate into enhanced downstream CNS-OT prediction performance. By recovering SCR components from EDA signals corrupted by severe MA and environmental noise, the model reconstructs cleaner signals while preserving both tonic and phasic components. This leads to improvements in accuracy, AUROC, and early prediction time of CNS-OT onset. These results suggest that improvements in denoising performance are directly associated with enhanced CNS-OT prediction performance, highlighting the importance of signal quality in reliable downstream clinical inference. Furthermore, the proposed lightweight denoising framework is not only suitable for resource-constrained deployment but also clinically meaningful for real-time early warning systems.

Despite these strengths, several limitations remain. First, the evaluation is conducted on a limited number of subjects, particularly in the UMAC (n = 13) and CNS-OT (n = 30) datasets, which may affect generalizability across broader populations. Second, although deployment feasibility is demonstrated through computational analysis, real-time implementation on embedded hardware was not directly validated. Future work will focus on large-scale validation across diverse populations and on-device implementation, including optimization techniques such as model quantization and hardware-aware design, to further assess real-world performance.

## VII. Conclusion

In this study, we propose a robust and deployable EDA denoising framework that integrates a hybrid CNN–Transformer (teacher) architecture with a lightweight depth wise separable CNN (student) trained via a KD strategy. The proposed approach enables robust EDA signal denoising under severe noise conditions while maintaining computational efficiency and preserving physiologically meaningful features (SCRs), making it suitable for deployment in wearable IoT devices.

The KD-based student model achieves substantial performance improvements over its baseline (the student model without KD), reducing MAE from 0.192 to 0.144 and RMSE from 0.243 to 0.185, while improving SNR from 9.828 dB to 12.08 dB, thereby narrowing the performance gap with the teacher model.

In real-world underwater conditions (UMAC dataset), the proposed method further enhances SCR reconstruction, reducing MAE from 2.809 to 0.215 under cvxEDA [37] and from 0.179 to 0.065 under ospEDA [38]. It also outperforms prior denoising methods based on filtering [39], wavelet transforms [15], and the U-Net [16] model. These results confirm the model's robustness in handling severe motion artifacts and environmental noise.

Notably, the improved signal quality translates into enhanced downstream clinical performance. The proposed method achieves the highest AUROC across both cvxEDA (0.806) and ospEDA (0.794) frameworks. In addition, under ospEDA, sensitivity (early prediction rate) improves from 0.550 (without denoising) to 0.767, demonstrating improved prediction capability and reliability for CNS-OT prediction.

In addition, the proposed student model reduces model size from 7.87 MB to 0.51 MB and FLOPs from 105.1M to 11.61M, while maintaining competitive inference latency (2.94 ms). These characteristics make the model well-suited for deployment on resource-constrained wearable IoT devices, such as MCUs with limited Flash memory (e.g., < 1 MB).

Overall, this work demonstrates that our lightweight EDA denoising framework can bridge the gap between signal reconstruction quality, clinical prediction performance, and practical deployment, enabling reliable physiological monitoring in challenging real-world environments.

## Acknowledgment

This work was supported by DHA23B-003. Fig. 2 was generated using Google Gemini (Gemini 3 Flash Image). During the preparation of this manuscript, the authors utilized ChatGPT to verify grammar and polish the language. Following the use of this service, the authors reviewed and edited the manuscript, and take full responsibility for its final content. The source code and associated models are publicly available at: [github.com/yongbin98/uEDA-Net](github.com/yongbin98/uEDA-Net). The UMAC dataset is available at: [github.com/jzizza/UMAC-Dataset](github.com/jzizza/UMAC-Dataset)

## References

[1] W. Boucsein, *Electrodermal Activity*. Springer Science & Business Media, 2012.

[2] M. S. Mahmud, H. Fang, and H. Wang, "An Integrated Wearable Sensor for Unobtrusive Continuous Measurement of Autonomic Nervous System," *IEEE Internet Things J.*, vol. 6, no. 1, pp. 1104–1113, Feb. 2019, doi: 10.1109/JIOT.2018.2868235.

[3] S. Aziz, G. Chetty, R. Goecke, and R. F. Rojas, "MDNet: A Lightweight Multi-Domain 1D-CNN for Embedded Pain Assessment Using EDA Signals," *IEEE Internet Things J.*, pp. 1–1, 2026, doi: 10.1109/JIOT.2026.3663683.

[4] A. Yazdinejad, H. Karimipour, and T. Halabi, "Toward Stress-Adaptive Cyber Defense: Cognitive–Physiological Synchronization in IoT Environments," *IEEE Internet Things J.*, vol. 13, no. 7, pp. 13832–13848, Apr. 2026, doi: 10.1109/JIOT.2026.3656466.

[5] Y. Kong and K. H. Chon, "Electrodermal activity in pain assessment and its clinical applications," *Appl. Phys. Rev.*, vol. 11, no. 3, p. 031316, Aug. 2024, doi: 10.1063/5.0200395.

[6] S. Vieluf *et al.*, "Twenty-four-hour patterns in electrodermal activity recordings of patients with and without epileptic seizures," *Epilepsia*, vol. 62, no. 4, pp. 960–972, 2021, doi: 10.1111/epi.16843.

[7] H. F. Posada-Quintero, C. S. Landon, N. M. Stavitzski, J. B. Dean, and K. H. Chon, "Seizures Caused by Exposure to Hyperbaric Oxygen in Rats Can Be Predicted by Early Changes in Electrodermal Activity," *Front. Physiol.*, vol. 12, Jan. 2022, doi: 10.3389/fphys.2021.767386.

[8] T. T. Wingelaar, P.-J. A. M. van Ooij, and R. A. van Hulst, "Oxygen Toxicity and Special Operations Forces Diving: Hidden and Dangerous," *Front. Psychol.*, vol. 8, Jul. 2017, doi: 10.3389/fpsyg.2017.01263.

[9] H. F. Posada–Quintero *et al.*, "Elevation of spectral components of electrodermal activity precedes central nervous system oxygen toxicity symptoms in divers," *Commun Med*, vol. 4, no. 1, p. 270, Dec. 2024, doi: 10.1038/s43856-024-00688-4.

[10] M.-B. Hossain *et al.*, "Prediction of central nervous system oxygen toxicity symptoms using electrodermal activity and machine learning," *Biocybernetics and Biomedical Engineering*, vol. 44, no. 2, pp. 304–311, Apr. 2024, doi: 10.1016/j.bbe.2024.03.004.

[11] Y. Zhang, M. Haghdan, and K. S. Xu, "Unsupervised motion artifact detection in wrist-measured electrodermal activity data," in *Proceedings of the 2017 ACM International Symposium on Wearable Computers*, in ISWC '17. New York, NY, USA: Association for Computing Machinery, Sep. 2017, pp. 54–57. doi: 10.1145/3123021.3123054.

[12] M.-B. Hossain, H. F. Posada-Quintero, Y. Kong, R. McNaboe, and K. H. Chon, "Automatic motion artifact detection in electrodermal activity data using machine learning," *Biomed. Signal Process. Control*, vol. 74, p. 103483, Apr. 2022, doi: 10.1016/j.bspc.2022.103483.

[13] Y. Kong, M. B. Hossain, A. Peitzsch, H. F. Posada-Quintero, and K. H. Chon, "Automatic motion artifact detection in electrodermal activity signals using 1D U-net architecture," *Comput. Biol. Med.*, vol. 182, p. 109139, Nov. 2024, doi: 10.1016/j.compbiomed.2024.109139.

[14] W. Chen, N. Jaques, S. Taylor, A. Sano, S. Fedor, and R. W. Picard, "Wavelet-based motion artifact removal for electrodermal activity," in *2015 37th Annual International Conference of the IEEE Engineering in Medicine and Biology Society (EMBC)*, Aug. 2015, pp. 6223–6226. doi: 10.1109/EMBC.2015.7319814.

[15] J. Shukla, M. Barreda-Ángeles, J. Oliver, and D. Puig, "Efficient wavelet-based artifact removal for electrodermal activity in real-world applications," *Biomed. Signal Process. Control*, vol. 42, pp. 45–52, Apr. 2018, doi: 10.1016/j.bspc.2018.01.009.

[16] M.-B. Hossain, H. F. Posada-Quintero, and K. H. Chon, "A Deep Convolutional Autoencoder for Automatic Motion Artifact Removal in Electrodermal Activity," *IEEE Trans. Biomed. Eng.*, vol. 69, no. 12, pp. 3601–3611, Dec. 2022, doi: 10.1109/TBME.2022.3174509.

[17] M.-B. Hossain, Y. Kong, H. F. Posada-Quintero, and K. H. Chon, "Comparison of Electrodermal Activity from Multiple Body Locations Based on Standard EDA Indices' Quality and Robustness against Motion Artifact," *Sensors*, vol. 22, no. 9, p. 3177, Jan. 2022, doi: 10.3390/s22093177.

[18] X. Shui *et al.*, "Bodily Electrodermal Representations for Affective Computing," *IEEE Trans. Affect. Comput.*, vol. 15, no. 3, pp. 1018–1025, Jul. 2024, doi: 10.1109/TAFFC.2023.3315973.

[19] Y. Kong *et al.*, "Sex differences in autonomic functions and cognitive performance during cold-air exposure and cold-water partial immersion," *Front. Physiol.*, vol. 15, Oct. 2024, doi: 10.3389/fphys.2024.1463784.

[20] R. W. Picard, S. Fedor, and Y. Ayzenberg, "Multiple Arousal Theory and Daily-Life Electrodermal Activity Asymmetry," *Emotion Review*, vol. 8, no. 1, pp. 62–75, Jan. 2016, doi: 10.1177/1754073914565517.

[21] J. Lin, W.-M. Chen, Y. Lin, J. Cohn, C. Gan, and S. Han, "MCUNet: Tiny Deep Learning on IoT Devices," in *Advances in Neural Information Processing Systems*, 2020, pp. 11711–11722.

[22] G. Hinton, O. Vinyals, and J. Dean, "Distilling the Knowledge in a Neural Network," Mar. 09, 2015, *arXiv*: 1503.02531. doi: 10.48550/arXiv.1503.02531.

[23] E. Perez, F. Strub, H. de Vries, V. Dumoulin, and A. Courville, "FiLM: Visual Reasoning with a General Conditioning Layer,"

*Proceedings of the AAAI Conference on Artificial Intelligence*, vol. 32, no. 1, Apr. 2018, doi: 10.1609/aaai.v32i1.11671.

[24] D. R. Bach, G. Flandin, K. J. Friston, and R. J. Dolan, "Modelling event-related skin conductance responses," *International Journal of Psychophysiology*, vol. 75, no. 3, pp. 349–356, Mar. 2010, doi: 10.1016/j.ijpsycho.2010.01.005.

[25] D. R. Bach, G. Flandin, K. J. Friston, and R. J. Dolan, "Time-series analysis for rapid event-related skin conductance responses," *Journal of Neuroscience Methods*, vol. 184, no. 2, pp. 224–234, Nov. 2009, doi: 10.1016/j.jneumeth.2009.08.005.

[26] D. R. Bach, J. Daunizeau, K. J. Friston, and R. J. Dolan, "Dynamic causal modelling of anticipatory skin conductance responses," *Biological Psychology*, vol. 85, no. 1, pp. 163–170, Sep. 2010, doi: 10.1016/j.biopsycho.2010.06.007.

[27] M. Staib, G. Castegnetti, and D. R. Bach, "Optimising a model-based approach to inferring fear learning from skin conductance responses," *Journal of Neuroscience Methods*, vol. 255, pp. 131–138, Nov. 2015, doi: 10.1016/j.jneumeth.2015.08.009.

[28] G. Castegnetti, A. Tzovara, M. Staib, S. Gerster, and D. R. Bach, "Assessing fear learning via conditioned respiratory amplitude responses," *Psychophysiology*, vol. 54, no. 2, pp. 215–223, 2017, doi: 10.1111/psyp.12778.

[29] S. Khemka, A. Tzovara, S. Gerster, B. B. Quednow, and D. R. Bach, "Modeling startle eyeblink electromyogram to assess fear learning," *Psychophysiology*, vol. 54, no. 2, pp. 204–214, 2017, doi: 10.1111/psyp.12775.

[30] C. W. Korn, M. Staib, A. Tzovara, G. Castegnetti, and D. R. Bach, "A pupil size response model to assess fear learning," *Psychophysiology*, vol. 54, no. 3, pp. 330–343, 2017, doi: 10.1111/psyp.12801.

[31] D. R. Bach, E. Seifritz, and R. J. Dolan, "Temporally Unpredictable Sounds Exert a Context-Dependent Influence on Evaluation of Unrelated Images," *PLOS ONE*, vol. 10, no. 6, p. e0131065, Jun. 2015, doi: 10.1371/journal.pone.0131065.

[32] D. R. Bach, "A head-to-head comparison of SCRalyze and Ledalab, two model-based methods for skin conductance analysis," *Biological Psychology*, vol. 103, pp. 63–68, Dec. 2014, doi: 10.1016/j.biopsycho.2014.08.006.

[33] D. Kang *et al.*, "Mechanically Robust Superhydrophobic Coatings via Dual-Step Deposition for Electrodermal Activity (EDA) Electrodes Immersible in Saltwater," Apr. 24, 2026, *Social Science Research Network, Rochester, NY*: 6641936. doi: 10.2139/ssrn.6641936.

[34] J. Chen *et al.*, "TransUNet: Transformers Make Strong Encoders for Medical Image Segmentation," Feb. 08, 2021, *arXiv*: 2102.04306. doi: 10.48550/arXiv.2102.04306.

[35] Y. Lee and K. H. Chon, "Atrial Fibrillation Prediction Using a Lightweight Temporal Convolutional and Selective State Space Architecture," Aug. 26, 2025, *arXiv*: 2508.19361. doi: 10.48550/arXiv.2508.19361.

[36] A. G. Howard *et al.*, "MobileNets: Efficient Convolutional Neural Networks for Mobile Vision Applications," Apr. 17, 2017, *arXiv*: 1704.04861. doi: 10.48550/arXiv.1704.04861.

[37] A. Greco, G. Valenza, A. Lanata, E. P. Scilingo, and L. Citi, "cvxEDA: A Convex Optimization Approach to Electrodermal Activity Processing," *IEEE Trans. Biomed. Eng.*, vol. 63, no. 4, pp. 797–804, Apr. 2016, doi: 10.1109/TBME.2015.2474131.

[38] Y. Lee, Y. Kong, and K. H. Chon, "ospEDA: Orthogonal Subspace Projection for Electrodermal Activity Decomposition," Apr. 08, 2026, *arXiv*: 2604.07521. doi: 10.48550/arXiv.2604.07521.

[39] J. Hernandez, R. R. Morris, and R. W. Picard, "Call Center Stress Recognition with Person-Specific Models," in *Affective Computing and Intelligent Interaction*, S. D'Mello, A. Graesser, B. Schuller, and J.-C. Martin, Eds., Berlin, Heidelberg: Springer, 2011, pp. 125–134. doi: 10.1007/978-3-642-24600-5_16.